\documentclass[12pt]{iopart}
\usepackage{graphicx}
\usepackage{pifont}
\usepackage{subfigure}
\usepackage{bm}
\usepackage[colorlinks,urlcolor=blue,anchorcolor=blue,linkcolor=blue,citecolor=blue,breaklinks=true]{hyperref}
\begin{document}

\title[]{Characterizations of topological superconductors: Chern numbers, edge states and Majorana zero modes}

\author{Xiao-Ping Liu$^1$, Yuan Zhou$^{1}$, Yi-Fei Wang$^2$
and Chang-De Gong$^{2,1,3}$}

\address{$^1$National Laboratory of Solid State Microstructures and
Department of Physics, Nanjing University, Nanjing 210093, China}
\address{$^2$Center for Statistical and Theoretical Condensed
Matter Physics, and Department of Physics, Zhejiang Normal
University, Jinhua 321004, China}
\address{$^3$Collaborative Innovation Center of Advanced Microstructures, Nanjing University, Nanjing 210093, China
}

%\ead{zhouyuan@nju.edu.cn}

%\begin{indented}
%\item[]June 6, 2017
%\end{indented}

\begin{abstract}
The topological properties in topological superconductors are usually characterized by the bulk Chern numbers, edge-state spectra, and Majorana zero modes. Whether they are equivalent or inequivalent is not well understood. Here, we investigate this issue with focus on a checkerboard-lattice model combining the Chern insulator and chiral $p$-wave superconductivity. Multiple topologically superconducting phases with Chern numbers up to $\mathcal{N}=4$ are produced. We explicitly demonstrate the mismatch between the Chern numbers, edge states and Majorana zero modes in this two-dimensional topological-superconductor model. The intrinsic reason is that some edge states in the superconducting phases inherited from the Chern-insulator phase are not protected by the particle-hole symmetry. We further check the mismatches in vortex states. Our results therefore clarify these different but complementary topological features and suggest that further considerations are required to characterize various topological superconductors.
\end{abstract}

\maketitle

\section{Introduction}

Chiral topological superconductors (TSCs) have been arousing great interest in recent years~\cite{Qi-RMP2011} due to the potential applications in the topological quantum computation~\cite{Tewari-PRL2007,Nayak-RMP2008,Alicea-RPP2012,Alicea-NP2011}. Edges~\cite{Alicea-RPP2012,Alicea-NP2011} or vortex cores~\cite{Jackiw-NPB1981,Volovik-JETP1999,Ivanov-PRL2001,SunHH-PRL2016} of chiral TSCs hold the Majorana zero modes, with quasiparticles as their own antiparticles~\cite{Majorana-Book2006,Wilczek-NP2009}, carrying the non-Abelian statistics~\cite{Ivanov-PRL2001}. A straightforward way to generate the chiral TSCs is the spin-triplet $p$-wave superconductivity in the normal metal, where the Majorana zero modes locate near the edges~\cite{Read-PRB2000,Kitaev-PU2001,Potter-PRL2010,Niu-PRB2012,Russo-PRB2013}. Recently, the conventional $s$-wave superconductivity has been further utilized to generate TSCs via proximity to some materials~\cite{Lutchyn-PRL2010,Oreg-PRL2010,Mourik-Science2012,Das-NP2012,Deng-NL2012,Nadj-Science2014,Lee-Science2014,Fu-PRL2008,Sau-PRL2010,Rokhinson-NP2012,Xu-PRL2015,SunHH-PRL2016}. For example, some TSC states with Chern numbers $2N-1$ have been proposed to be realized near the quantum Hall or quantum anomalous Hall plateau transition from Chern numbers $N-1$ to $N$ by proximity to the $s$-wave superconductors~\cite{Qi-PRB2010,Wang-PRB2015,Wang-PRB2016}. Hence, question as to how the unconventional $p$-wave superconductivity affects the nontrivial topological states is naturally concerned about. We have proposed a spinless fermion model with chiral $p_{x}+ip_{y}$ pairing between the nearest neighbors in the checkerboard-lattice Chern insulator (CI), where rich TSC phases, including states with high Chern numbers up to $\mathcal{N}=3$, are induced by the harmonic trap potential~\cite{Liu-SR2016}.

On the other hand, how to characterize the topological properties of TSCs is also a widely concerned issue. In two-dimensional Chern-insulator lattice systems, the topological properties are well defined by the bulk-state Chern numbers~\cite{Thouless-PRL1982} or the winding numbers of chiral edge states~\cite{Hatsugai-PRL1993} in confined geometries. The Chern number is meaningless for the time-reversal invariant system. Consequently, a spin Chern number~\cite{ShengL-PRL2005,ShengD-PRL2006,Prodan-PRB2009} is introduced to describe the topological properties of quantum spin Hall insulators. In TSCs, both the Chern numbers and edge states have been utilized to characterize the topological properties. Another remarkable feature in TSCs is the existence of the Majorana zero modes. Whether these characterizations are equivalent or not has not been discussed in details for TSCs. A simple example is the $\mathcal{N}=2$ TSC state in the quantum anomalous Hall state by proximity to the conventional $s$-wave superconductor, where Majorana zero mode does not show up in the energy spectrum~\cite{Qi-PRB2010}.

Here, based on the checkerboard-lattice CI model with the chiral next-nearest-neighbor $p$-wave pairing, the Chern numbers, the edge-state spectra and the Majorana zero modes are analyzed in details. Rich TSC phases with Chern numbers $\mathcal{N}$ up to $4$ are obtained by adjusting the model parameters. We explicitly demonstrate the mismatches between the bulk Chern numbers, the edge states and the Majorana zero modes. The essential point is that part of the edge states originated from the CI is not preserved by the particle-hole symmetry, and therefore contributes to the Chern number but does not to the number of Majorana zero modes. Interestingly, both the edge states and Majorana zero modes exist in a special state with $\mathcal{N}=0$. We further check our statements in the vortex cores; similar mismatches are also discovered. Our results thus provide the detailed understanding of the topological properties in various TSCs and imply the complexity for assigning topological numbers to them.

\section{Model and formulation}
\label{Sec.2}
\begin{figure}[!htb]
\centering
\includegraphics[width=3.0in]{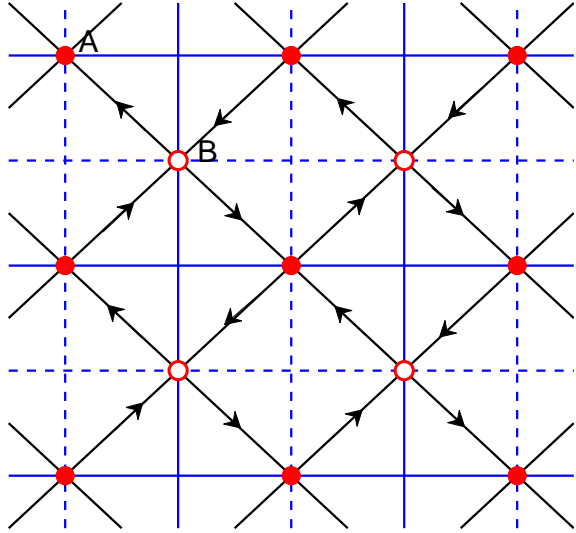}
\caption{(Color online). The checkerboard-lattice CI model. The sublattices are labeled by solid and hollow circles. The phase factor $\pm\phi$ on the nearest-neighbor hopping is denoted by the arrow. Blue solid and dash lines characterize the next-nearest-neighbor hoppings with different signs.}\label{f.1}
\end{figure}
We adopt a fermion model in checkerboard lattice to generate the CI as schematically illustrated in Fig.~\ref{f.1}~\cite{SunK-PRL2009,SunK-PRL2011,Liu-SR2016}. To avoid the complexity from the spin degrees of freedom, e.g., the spin Chern number and helical edge states, only spinless fermions are considered. The staggered fluxes are imposed on the plaquettes, divide the lattice into two sublattices $A$ and $B$, and induce additional phase factors $\pm\phi$ on the nearest-neighbor hoppings, which are essential to generate two topological bands with Chern numbers $\pm1$ and open an energy gap between them. On the other hand, the next-nearest-neighbor hoppings change signs along the different directions and between the different sublattices, and preserve the particle-hole symmetry of energy bands. The CI model is expressed as
\begin{equation}
\mathcal{H}_{\rm CI}=-t\sum_{\langle i,j\rangle}(e^{i\phi_{ij}}c_{i}^{\dagger}c_{j}+ {\rm H.c.})-\sum_{\langle\langle i,j\rangle\rangle}(t^{\prime}_{ij}c_{i}^{\dagger}c_{j}+ {\rm H.c.}).
\label{E.1}
\end{equation}
Here, $c_{i}$ ($c_{i}^{\dagger}$) annihilates (creates) a spinless fermion on the site $i$. The staggered fluxes are superimposed in the plaquettes and induce an additional phase factor $\phi_{ij}=\pm\phi$ on the nearest-neighbor hopping $t$ with $\pm$ denoted by the arrow. $t^{\prime}_{ij}$ are the hopping parameters for the next-nearest neighbors with different sign (Fig.~\ref{f.1}). In numerics, $t=1$ is set as unit and $t^{\prime}=0.5$. When $\phi\neq0,\pi$ and at half-filling, the system becomes a CI with two well separated topological bands (See Appendix) carrying Chern numbers $+1$ and $-1$ respectively. Such a CI model is expected to be realizable in optical lattices with the recent technical developments for generating artificial gauge fields~\cite{Dalibard-RMP2011,Aidelsburger-PRL2011,Jimenez-Garcia-PRL2012,Struck-PRL2012}.

We further include the chiral $p_{x}+ip_{y}$ superconducting pairing between the same sublattice in this CI model, i.e.,
\begin{equation}
\mathcal{H}_{\rm SC}=\sum_{\langle\langle i,j\rangle\rangle}(\Delta_{ij}c_{i}^{\dagger}c_{j}^{\dagger}+ {\rm H.c.}),
\label{E.1}
\end{equation}
where $\Delta_{ij}=-V\langle c_{j}c_{i}\rangle$ is the pairing potential
with $V$ strength of the attractive interaction.

By the Fourier transformation, the Bogoliubov-de Gennes Hamiltonian can be rewritten in momentum space as
\begin{eqnarray}
    H=\frac{1}{2}\sum_{\bm{k}}\left(
                              \begin{array}{ccc}
                              c_{\bm{k}}^{\dagger} &
                              c_{\bm{-k}}
                              \end{array}
                              \right)
                              \left(
                              \begin{array}{ccc}
                              H_{0}(\bm{k}) & \Delta(\bm{k})\\
                              \Delta^{\dagger}(\bm{k}) & -H_{0}^{T}(-\bm{k})
                              \end{array}
                              \right)
                              \left(
                              \begin{array}{ccc}
                              c_{\bm{k}}\\
                              c_{\bm{-k}}^{\dagger}
                              \end{array}
                              \right)
\label{E.2}
\end{eqnarray}
where $\left(c_{\bm{k}}^{\dagger}, c_{\bm{-k}}\right) = \left( a_{
\bm{k}}^{\dagger}, b_{\bm{k}}^{\dagger}, a_{\bm{-k}}, b_{\bm{-k}} \right)$. $H_{0}(\bm{k})=\epsilon_{k}^{z}\sigma_{z}+\epsilon_{k}^{x}\sigma_{x}+\epsilon_{k}^{y}\sigma_{y}-\mu
\mathcal{I}$ with $\epsilon_{k}^{x}=-4t\cos
\phi\cos\frac{k_{x}}{2}\cos\frac{k_{y}}{2}$, $\epsilon_{k}^{y}=-4t\sin
\phi\sin\frac{k_{x}}{2}\sin\frac{k_{y}}{2}$, and
$\epsilon_{k}^{z}=-2t^{\prime}(\cos k_{x}-\cos k_{y})$. $\sigma$, and
$\mathcal{I}$ is the Pauli matrix, and the unit matrix respectively. $\mu$ is the chemical potential to preserve the particle number conservation. The chiral
$p_{x}+ip_{y}$ pairing of the next-nearest neighbors is of the form
$\Delta(\bm{k})=2\Delta[-\sin k_{y}+i\sin k_{x}]\mathcal{I}$
with $\Delta$ the superconducting order parameter.  %With $\Delta(\bm{k})$=0, critical conditions of TQPTs can be ascertained
%and are independent of the superconducting order parameter $\Delta$ which are different from those of cases with
%nearest-neighbor superconducting pairing in our previous work~\cite{Liu-SR2016}.

To study Majorana zero modes, we solve the Bogoliubov-de Gennes (BdG) Hamiltonian in real space self-consistently,
\begin{eqnarray}
    \sum_{j}\left(
              \begin{array}{ccc}
                 H_{ij} & \Delta_{ij}\\
                \Delta_{ji}^{*} & -H_{ij}^{T} \\
              \end{array}
            \right)
            \left(
              \begin{array}{ccc}
                 u_{j}^{n}\\
                 v_{j}^{n}\\
              \end{array}
            \right)
            =E_{n}\left(
              \begin{array}{ccc}
                 u_{i}^{n}\\
                 v_{i}^{n}\\
              \end{array}
            \right)
\end{eqnarray}
where $H_{ij}=-te^{i\phi_{ij}}\delta_{i+\tau,j}-t^{\prime}_{ij}\delta_{i+\tau',j}-\mu\delta_{ij}$
is the single-particle Hamiltonian with $\tau$ and $\tau'$ vectors
linked by the nearest-neighbor and next-nearest-neighbor sites. $\left(u_{i}^{n},
v_{i}^{n}\right)^{T}$ is the quasiparticle wave function
corresponding to the eigenvalue $E_{n}$. Due to the particle-hole
symmetry of the BdG equations, the wave vector $\left(v_{i}^{n*},
u_{i}^{n*}\right)^{T}$ is also an eigenvector corresponding to
eigenvalue $-E_{n}$. The superconducting order parameter can be
determined self-consistently by
\begin{eqnarray}
    \Delta_{ij}&=&\frac{V}{2}\sum_{n}u_{i}^{n}v_{j}^{n*}\tanh(E_{n}/2k_{\rm B}T).
\end{eqnarray}

\section{Topological properties of TSCs}
\label{Sec.3}
\begin{figure}[!htb] \centering\subfigure[]{
\label{fig:subfig:a}
\hspace{-0.5in}\includegraphics[width=3.5in]{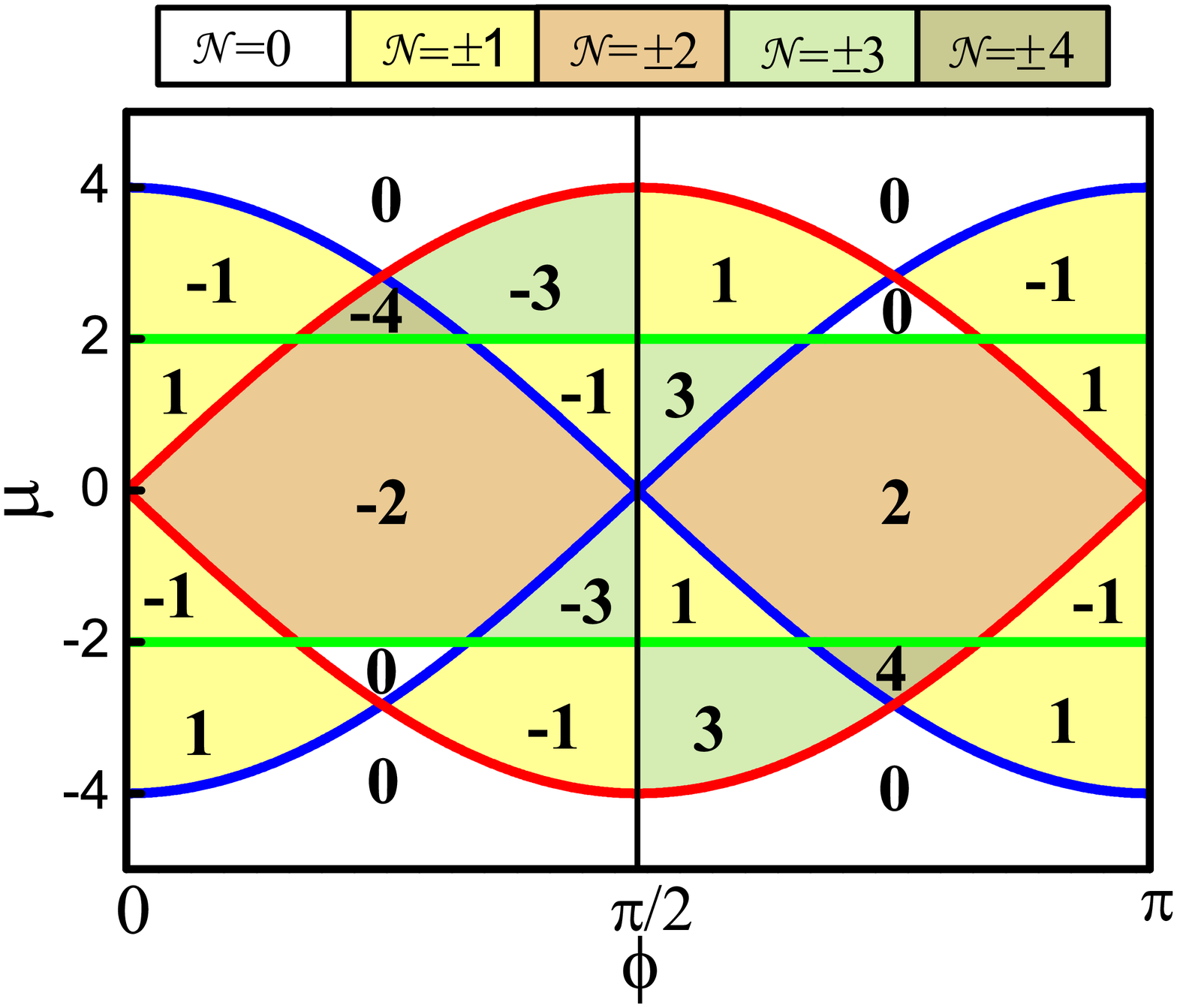}}
\subfigure[]{ \label{fig:subifg:b}
\hspace{-0.5in}\includegraphics[width=3.5in]{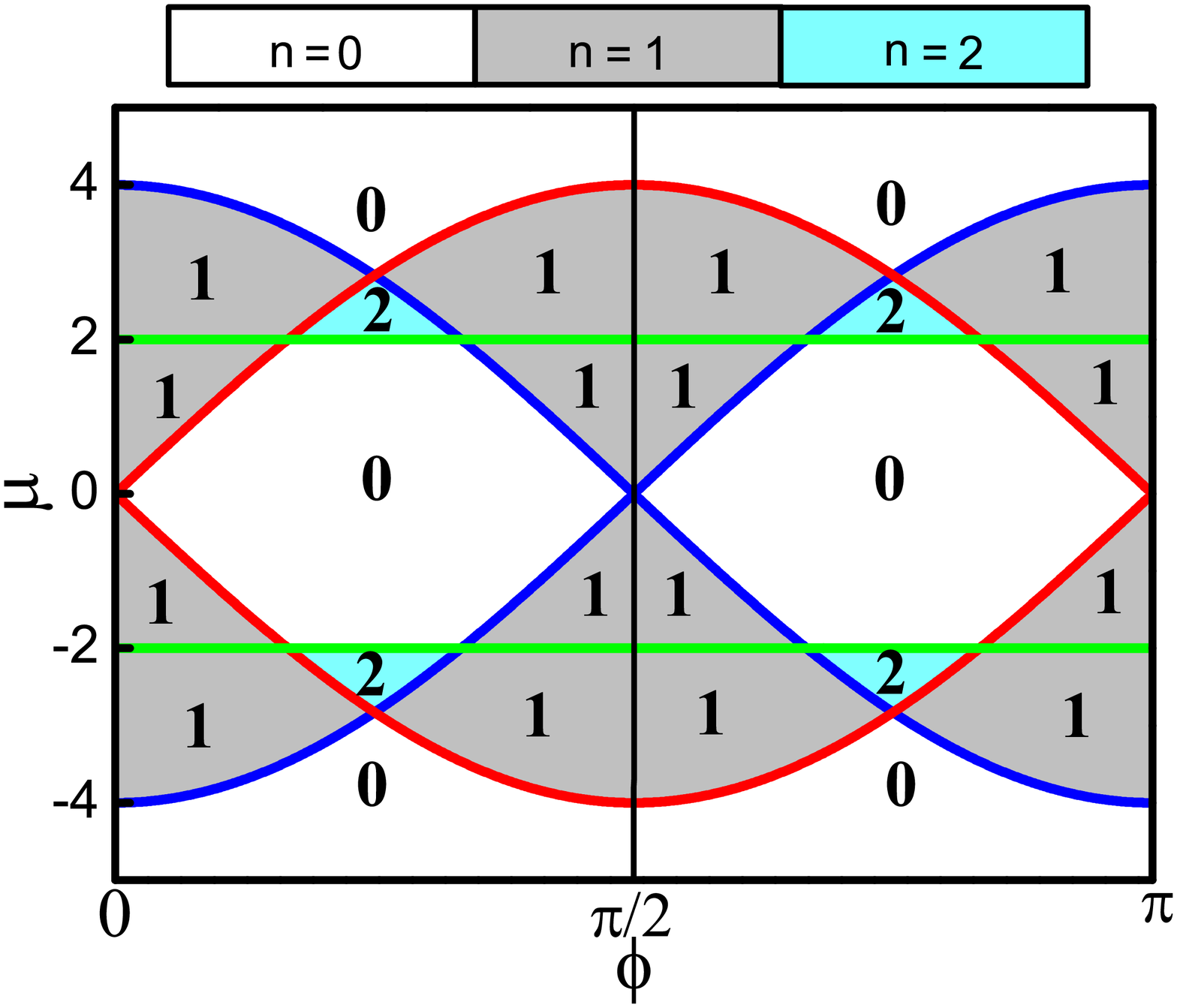}}
\caption{(Color online). Phase diagram in $\phi-\mu$ parameter space with the pairing parameter $\Delta=0.1$. Blue, red and green lines mark the phase boundaries $\mu=\pm4\cos\phi$, $\mu=\pm4\sin\phi$ and $\mu=\pm4t^{\prime}$, respectively. (a) Phases are characterized by Chern numbers $\mathcal{N}$. (b) Phases are characterized by the number of Majorana zero modes.}\label{fig:subfig}\label{f.2}
\end{figure}

The two-dimensional BdG Hamiltonian with broken time-reversal symmetry belongs to the topological class D, which is characterized by an integer number\cite{Altland-PRB1997,Schnyder-PRB2008}. The Chern number in an individual band is not well defined when the two lower BdG bands overlap in some cases. We use the sum of the two lower BdG bands $\mathcal{N}=C_{1}+C_{2}$ with $C_{1}$ and $C_{2}$ the Chern number for the lower two bands to characterize the topological properties of TSCs \cite{Qi-PRB2010,Liu-SR2016}. The topological phase boundaries can be analytically determined from the quasiparticle dispersion by the conditions at which the two middle BdG bands touch and re-open. Such considerations yield the phase boundary $\mu=\pm 4t\cos \phi$ with a single Dirac point at $(0,0)$, $\mu=\pm 4t^{\prime}$ with dual Dirac points at $(\pi,0)$ and $(0,\pi)$, and $\mu=\pm 4t\sin\phi$ with single touching point at $(\pi,\pi)$. Compared with the nearest-neighbor pairing between the different sublattices \cite{Liu-SR2016}, the condition $\mu=\pm 4t^{\prime}$ is independent of the superconducting order parameter since $\Delta_{k}=0$ at these Dirac points. All topological phases are summarized in Fig.~\ref{f.2}(a). Rich topological quantum phases with the Chern numbers up to $\mathcal{N}=4$ are obtained by changing the staggered-flux parameter $\phi$ and the chemical potential $\mu$.

\begin{figure}[!htb] \centering{
\hspace{-1.5in}\includegraphics[width=0.7\textwidth]{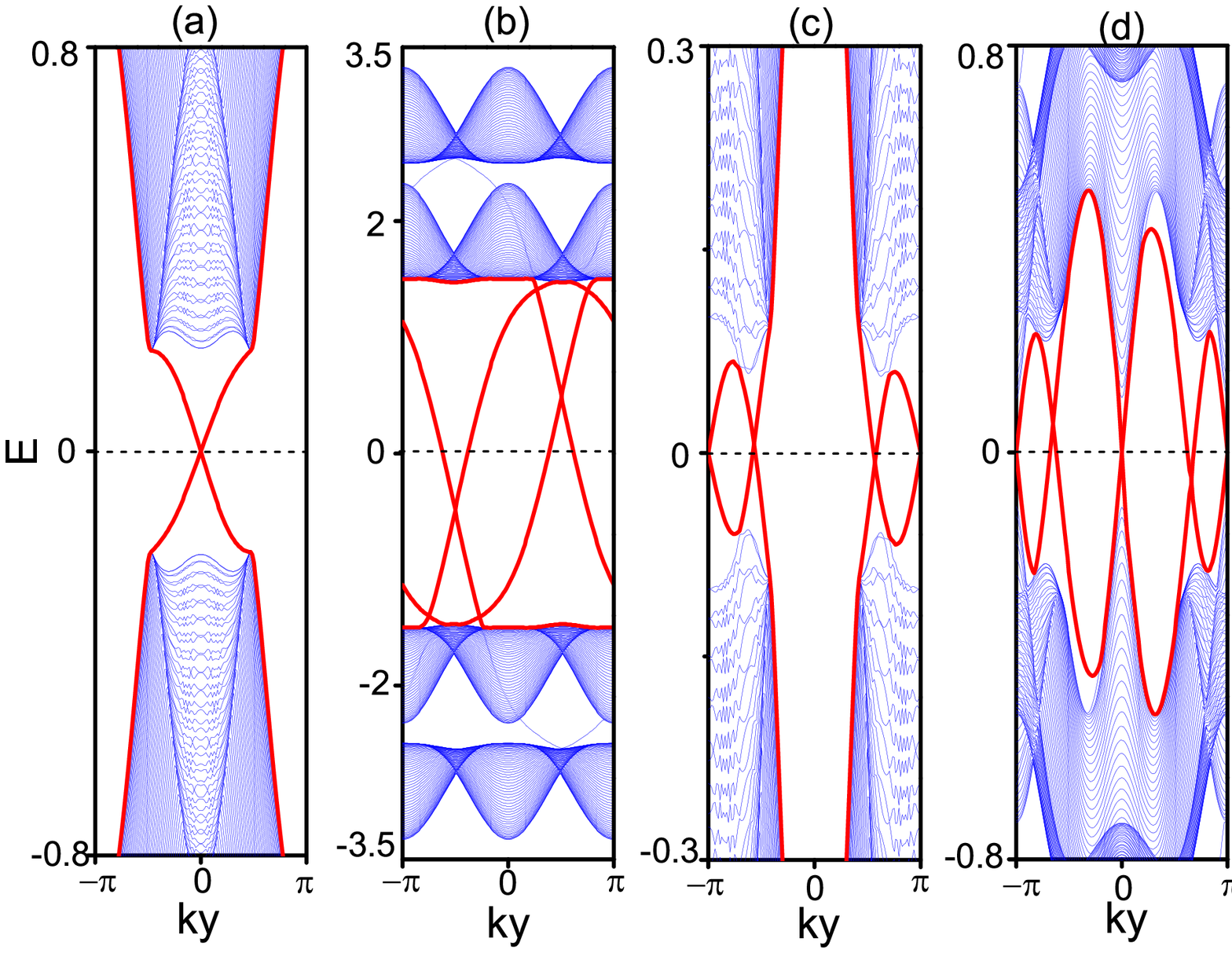}}
{
\includegraphics[width=\textwidth]{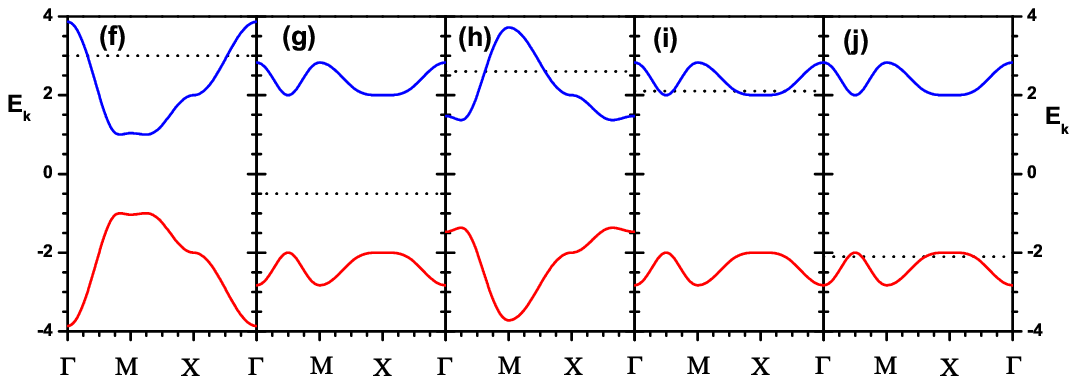}}
\caption{(Color online). Upper panels are the energy spectra of different TSCs on cylindrical geometry, the edge states are highlighted by red. From left to right, the parameters are: (a) $\phi=\pi/12$, $\mu=3$, $\Delta=0.1$, $\mathcal{N}=-1$. (b) $\phi=\pi/4$, $\mu=-0.5$, $\Delta=0.1$, $\mathcal{N}=-2$. (c) $\phi=0.38\pi$, $\mu=2.6$, $\Delta=0.1$, $\mathcal{N}=-3$. (d) $\phi=\pi/4$, $\mu=2.1$, $\Delta=0.4$, $\mathcal{N}=-4$. (e) $\phi=\pi/4$, $\mu=-2.1$, $\Delta=0.4$, $\mathcal{N}=0$. Lower panels are the corresponding CI bands (without superconductivity), Fermi energy is denoted by dotted line.}\label{f.3}
\end{figure}

We then turn to the edge states in TSCs, which are calculated for a cylindrical system with the open boundary condition along the $x$-direction and the period boundary condition along the $y$-direction~\cite{Potter-PRL2010}. For the $\mathcal{N}=1$ TSC state [Fig.~\ref{f.3}(a)], a pair of edge states propagate along the opposite boundaries, corresponding to the Chern number. As analyzed in the Appendix, the next-nearest-neighbor pairing turns to be the intra- and inter-CI band pairings after the rotational transformation (See Appendix); the former is protected by the particle-hole symmetry while the latter is not. Here, the pairing in the edge state is the intra-CI band pairing; the Dirac point at $k_{y}=0$ therefore locates at the zero energy. This TSC state is similar to the $\mathcal{N}=1$ TSC phase suggested by Read et al. with the chiral $p$-wave superconducting pairing in the trivial normal state~\cite{Read-PRB2000} and by Qi et al. with the quantum anomalous Hall state in proximity to a conventional $s$-wave superconductor \cite{Qi-PRB2010}, where the particle-hole symmetry preserves. In contrast, two pairs of edge states propagate along the edges in the $\mathcal{N}=2$ TSC state. However, the pairing in the edge states comes from the inter-CI band. The Dirac points deviate from zero energy [Fig.~\ref{f.3}(b)]. This is also consistent with the $\mathcal{N}=2$ TSC state found previously for the quantum anomalous Hall system in proximity to a $s$-wave superconductor, where two identical chiral Majorana edge modes originated from the CI are suggested~\cite{Qi-PRB2010}.

The mixed intra- and inter-CI band superconducting pairing in the edge states are found in the cases with high Chern numbers as shown in Fig.~\ref{f.3} (c) and (d). One Dirac point at $k_{y}=\pi$ from the intra-CI band pairing locates at zero energy preserved by particle-hole symmetry, while the rest two from the inter-CI band pairing locate above or below the zero energy in the $\mathcal{N}=3$ TSC states. Similarly, two Dirac points at $k_{y}=0$ and $k_{y}=\pi$ from the intra-CI band pairing locate at zero energy in the $\mathcal{N}=4$ TSC states. Therefore, the number of edge states, with two types of Dirac points, and the Chern number correspond one by one.

Interestingly, there are several branches of edge states and two Dirac points at $k_{y}=0$ and $k_{y}=\pi$ from the intra-CI band pairing (Fig.~\ref{f.3} (e)) in a special $\mathcal{N}=0$ state. This state is therefore also a nontrivial TSC state though with a trivial Chern number. It seems that the numbers of the edge states and Chern number are inconsistent in this case. A possible explanation is that the edge states propagate along the same boundary in the opposite directions, resulting in zero net current along the boundaries.

We notice that the band structure seems to be antisymmetric about $k_{y}$ for $\mathcal{N}=2$, $3$, $4$, and $0$ TSCs in the upper panels of Fig.~\ref{f.3}. This is a direct consequence of particle-hole symmetry in cylindrical geometry as discussed in Appendix \emph{II}. The particle-hole symmetry is guaranteed by $\mathcal{P}H(k_{y})\mathcal{P}^{-1}=-H(-k_{y})$, where the operator for particle-hole transformation is $\mathcal{P}=\sigma_{x}\otimes\mathcal{I}K$ with $\sigma$ the Pauli matrix, $\mathcal{I}$ the $2M\times 2M$ identity matrix ($M$ the number of sublattice along the $x$-direction) and K the complex conjugation. In this sense, the Majorana zero modes can be only found at $k_{y}=0$ or $k_{y}=\pi$, which is the exact result in the upper panels of Fig.~\ref{f.3}.

To better understand the particle-hole symmetry in respective topological state, we plot the corresponding CI bands in the lower panels of Fig.~\ref{f.3}. In $\mathcal{N}=-1$ TSC state, the Fermi energy cuts through the upper CI-band near the Dirac point $\Gamma=(0,0)$. Therefore, the intra-CI band pairing contributes a pair of zero modes when the open condition is applied on $x$-direction (Fig.~\ref{f.3}(f)). In contrast, the Fermi energy cuts neither upper nor lower CI band in $\mathcal{N}=-2$ TSC state (Fig.~\ref{f.3}(g)). The pairing is therefore inter-CI band pairing. The edge states in Fig.~\ref{f.3}(b) inherited from the CI phase are not protected by particle-hole symmetry. Similar analysis can be also applied in high Chern number TSC states. The Fermi energy cuts through the upper CI band near the touch point $M={\pi,\pi}$ in $\mathcal{N}=-3$ TSC state. The pairing is therefore the intra-CI band pairing and contribute a pair of zero modes. In $\mathcal{N}=-4$ TSC state, the Fermi energy cutting through the upper CI band around $\Gamma=(0,0)$ and $M=(\pi,\pi)$, yielding two pairs of zero modes. $\mathcal{N}=0$ TSC state is similar to $\mathcal{N}=-4$ TSC state but with Fermi energy cutting through the lower CI band.

\begin{figure*}[tbp]{\centering
  \includegraphics[width=0.7\textwidth]{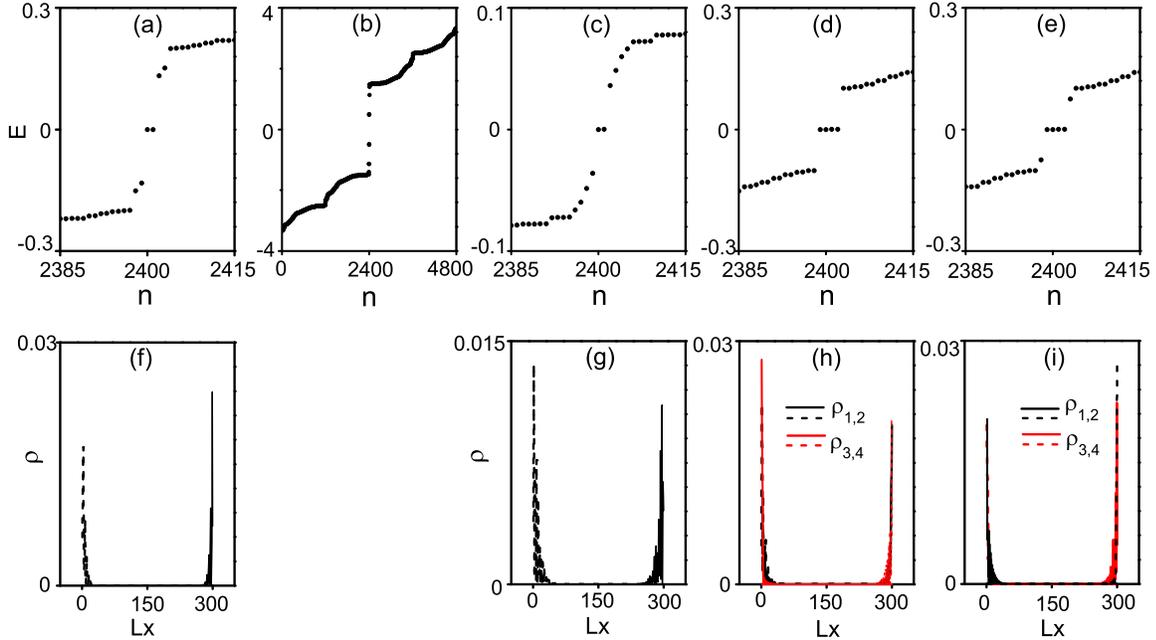}}
  \caption{(Color online). Energy spectra and distributions of the Majorana zero modes in real space. The calculations are performed on the lattice $L_{x}\times L_{y}=300\times8$ with open boundary condition along $x$-direction and periodic boundary condition along $y$-direction. The upper panels are the energy spectra for Chern number (a) $\mathcal{N}=-1$, (b) $\mathcal{N}=-2$, (c) $\mathcal{N}=-3$, (d) $\mathcal{N}=-4$ and (e) $\mathcal{N}=0$. The adopted parameters are the same as in Fig.~\ref{f.3}. Low panels are the corresponding distributions of Majorana zero modes. Note, $\mathcal{N}=-2$ TSC is a trivial state without Majorana zero modes.}\label{f.4}
\end{figure*}

The Majorana zero modes are now apparent from the above analysis. We check the zero modes by solving the BdG equations in real space. The corresponding energy spectra are shown in Fig.~\ref{f.4} (a)-(e). The numbers of the pairs of the Majorana zero modes are $1$, $0$, $1$, and $2$ for the $\mathcal{N}=1$, $2$, $3$, and $4$ TSC states, and the special case with the $\mathcal{N}=0$ state has two pairs of Majorana zero modes. These features are concluded in the phase diagram Fig.~\ref{f.2} (b), which exhibits evident particle-hole symmetry since the Majorana zero modes are protected by this symmetry. We have also checked the wave functions in the respective TSC states. The wave function has equivalent combination of the particles and holes at the Dirac points with zero energy, manifesting the particle-hole symmetric nature of Majorana zero modes again. In comparison, the particles and holes admixture is inequivalent at the Dirac points with non-zero energy. Those Majorana zero modes are well separated and locate on the boundaries [Fig.~\ref{f.4} (f)-(i)] as revealed before ~\cite{Liu-SR2016}. As far as the Majorana zero modes indicate, the state with $\mathcal{N}=0$, and $2$, is a TSC state, and a trivial TSC state \cite{Qi-PRB2010}, respectively. Anyway, the Chern numbers and Majorana zero modes may be mismatched with each other. The essential reason is that some of pairings in the edge states come from the inter-CI band pairing, which is not protected by the particle-hole symmetry.

Our study provides a detailed comparison among the Chern numbers, edge states, and Majorana zero modes, which are frequently used to characterize the topological nature of TSCs in various literatures. The Chern number describes the global topological property including either particle-hole symmetry protected or unprotected edge states in cylinder, whereas the Majorana zero modes contain only the former. The Chern number may be invalid to characterize the topological properties of TSCs in multi-band systems. Further considerations on the Chern numbers in TSCs are therefore necessary.

\section{Vortex states of TSCs}
Vortices holding Majorana zero modes might provide a promising platform for realizing non-Abelian statistics and topological quantum computation~\cite{Nayak-RMP2008,Alicea-RPP2012,Alicea-NP2011}. We will further check the relation between the Chern numbers and Majorana zero modes in the vortex states in this section. A uniform magnetic field is imposed on the periodic lattice with fixed size $56\times28$, where two vortices are contained by adjusting the magnitude of the magnetic field. Results are obtained by self-consistent calculations of the BdG equations in real space.

\begin{figure*}[tbp]
\centering
\includegraphics[width=0.8\textwidth]{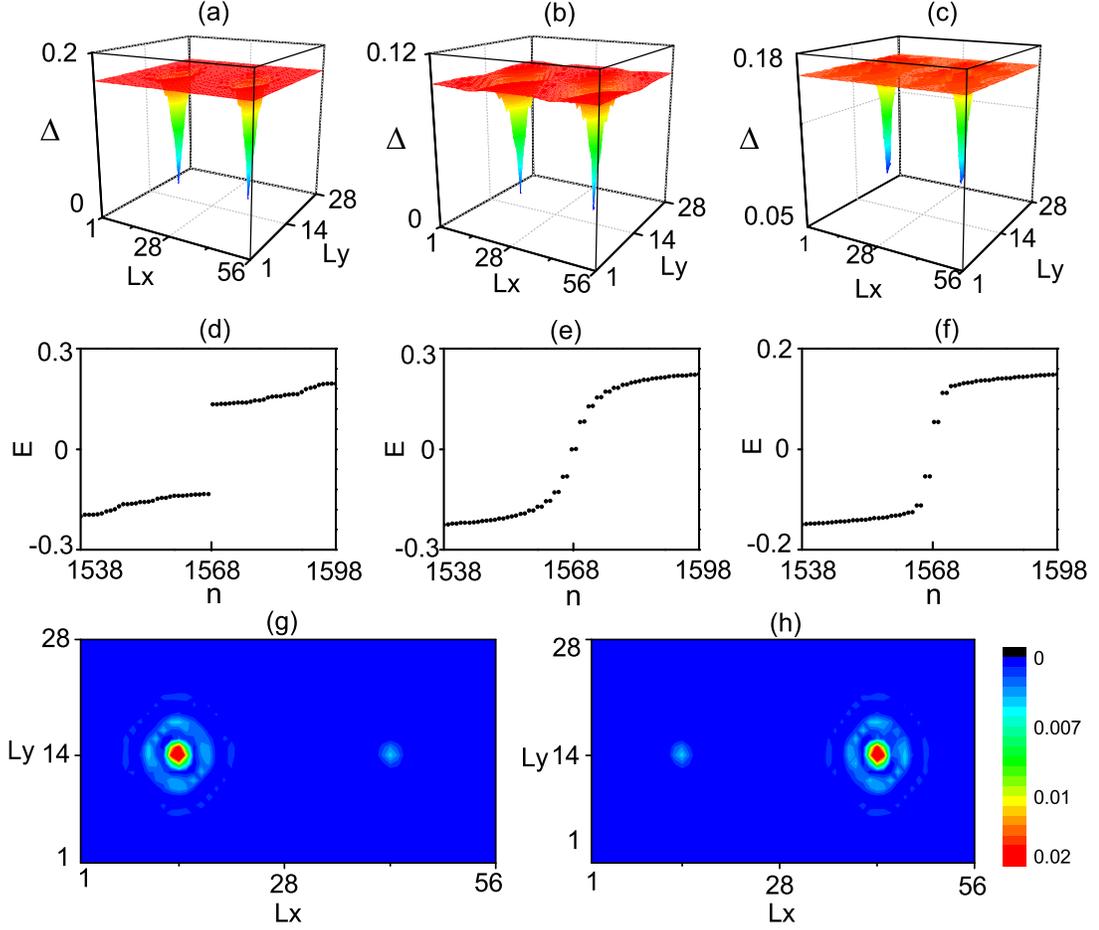}
\caption{(Color online). The superconducting order parameters and low-energy spectra of the vortex states. (a) and (d) are for $\mathcal{N}=2$ TSC with parameters $\phi=\pi/12$, $\mu=-0.9$, and $V=3$; (b) and (e) are for $\mathcal{N}=-3$ TSC with parameters $\phi=5\pi/12$, $\mu=3$, $V=2.4$; and (c) and (f) are for $\mathcal{N}=-4$ TSC with parameters $\phi=0.2\pi$, $\mu=2.2$, $V=2$. (g) and (h) display the distributions of Majorana zero modes in $\mathcal{N}=-3$ TSC with the same parameters in (b).}\label{f.5}
\end{figure*}

While the vortex state in the TSC with $\mathcal{N}=1$ has been investigated previously \cite{Read-PRB2000}, we here show the results for the TSCs with higher Chern numbers. The case with $\mathcal{N}=2$ is shown in Fig.~\ref{f.5} (a) and (d). The superconductivity is destroyed in the vortex core. However, no Majorana zero mode in the low-energy spectrum is found. This is consistent with our statement that $\mathcal{N}=2$ is a trivial TSC state as discussed in the section~\ref{Sec.3}, manifesting the mismatch between the Chern number and Majorana zero modes. In comparison, one pair of Majorana zero modes are found in $\mathcal{N}=3$ TSC state [Fig.~\ref{f.5} (b) and (e)], in agreement with the results in cylindrical geometry. The Majorana zero modes are well separated and mainly distribute in the vortex core where phase boundary locates. In fact, the distribution of Majorana zero mode also shows Friedel-like oscillations as shown in Fig.~\ref{f.5} (g) and (h), similar to the modulation induced by the impurity or vortex in two-dimensional lattice \cite{ZhuJX-PRL2002,ChenY-PRL2004}. Therefore, slight overlap between the Majorana zero modes can be found, consisting with previous study~\cite{ZhouT-PRB2013}. Unexpectedly, we have not found signatures of Majorana zero modes in the vortex states of the TSC with $\mathcal{N}=0$, and $4$ (Fig.~\ref{f.5} (c) and (f)). The possible reason is that the mixing between two zero modes in the vortex core can lead to energy splitting and therefore destroy the zero modes.

\section{Summary and discussion}
In summary, the topological properties in TSC have been studied in a checkerboard-lattice CI model together with the chiral $p$-wave superconductivity. Rich topological quantum phases have been obtained with Chern numbers up to $\mathcal{N}=4$. The TSC states with Chern numbers $\mathcal{N}=1$, $2$, $3$ and $4$  respectively hold $1$, $0$, $1$, and $2$ pairs of Majorana zero modes in cylindrical confinement, as well as a special $\mathcal{N}=0$ TSC state but with two pairs of Majorana zero modes at edges. Our results therefore imply that the Chern numbers and Majorana zero modes are not always matched with each other. The essence is that Chern number contains both the particle-hole symmetrically protected and unprotected edge states, whereas the Majorana zero modes are always protected by the particle-hole symmetry. We further check these statements in the vortex states of TSCs. Similar mismatches between the Chern numbers and the Majorana zero modes have also been revealed.

The present results strongly suggest that the Chern number, well corresponding to the BdG quasiparticle current near the edges, may be not adequate to characterize the topological properties of TSCs. Similar to spin Chern number in quantum Hall system ~\cite{ShengD-PRL2006}, reconsidering or redefining the topological number in TSC is probably necessary, especially in the multi-band system. On the other hand, the failure of producing the Majorana zero modes in some TSC vortex states implies the vortex state is not always ready to generate non-Abelian statistics and therefore its application in the topological quantum computations should be cautious.

\ack
We thank H. Q. Lin for helpful discussions and suggestions. This work was supported by the National Nature Science Foundation of China under Grants No. 11374265 and 11274276, the Ministry of Science and Technology of China under Grant No. 2016YFA0300401. Y. Zhou acknowledges the financial support of CSC.

\section*{Appendix I: Pairing channel and particle-hole symmetry}
\label{Adix1}
The CI lattice model is expressed in momentum space as
\begin{equation}
\mathcal{H}_{\rm CI}=\sum_{k}c_{\bm{k}}^{\dagger}\mathcal{H}_{0}\left( k\right) c_{\bm{k}},
\end{equation}%
where $c_{\bm{k}}^{\dag}=(a_{k}^{\dag}, b_{k}^{\dag})$ with $a_{k}$ and $b_{k}$ the annihilation operator for sublattice $A$ and $B$. $\mathcal{H}_{0}(\bm{k})=\epsilon^{z}_{k}\sigma_{z}+\epsilon^{x}_{k}\sigma_{x}+\epsilon^{y}_{k}\sigma_{y}-\mu\mathcal{I}$ with $\sigma$, and $\mathcal{I}$, the Pauli, and identical matrix, respectively. Introducing the
rotational transformation %
\begin{eqnarray}
\left(
\begin{array}{c}
a_{k} \\
b_{k}%
\end{array}%
\right) =\left(
\begin{array}{cc}
\cos \theta _{k} & e^{-i\varphi _{k}}\sin \theta _{k} \\
-e^{i\varphi _{k}}\sin \theta _{k} & \cos \theta _{k}%
\end{array}%
\right) \left(
\begin{array}{c}
\alpha_{k} \\
\beta_{k}%
\end{array}%
\right)
\end{eqnarray}%
with $%
\xi _{k}=\sqrt{\left( \epsilon _{k}^{x}\right) ^{2}+\left( \epsilon
_{k}^{y}\right) ^{2}+\left( \epsilon _{k}^{z}\right) ^{2}}$, $\cos ^{2}\theta
_{k}=\frac{1}{2}\left( 1+\frac{\epsilon ^{z}_{k}}{\xi _{k}}\right) $, $\sin
2\theta _{k}=-\frac{\left\vert \epsilon ^{x}_{k}+i\epsilon ^{y}_{k}\right\vert }{\xi
_{k}}$, and $e^{i\varphi _{k}}=\frac{\epsilon ^{x}_{k}+i\epsilon ^{y}_{k}}{%
\left\vert \epsilon ^{x}_{k}+i\epsilon ^{y}_{k}\right\vert }$, the explicit form of $\epsilon^{x,y,z}_{k}$ has been shown in the main text.
The Hamiltonian can be diagonalized as%
\begin{equation}
\bar{\mathcal{H}}_{\rm CI}=\xi _{k}^{\alpha}\alpha_{k}^{\dagger}\alpha_{k}+\xi _{k}^{\beta}\beta%
_{k}^{\dagger}\beta_{k}.
\end{equation}%
Here $\xi _{k}^{\alpha}=\xi _{k}-\mu$, and $\xi _{k}^{\beta}=-\xi _{k}-\mu$ is the quasiparticle
dispersion of the upper, and lower CI band, respectively.

\bigskip
The superconducting term with $p_{x}+ip_{y}$ pairing between the next-nearest neighbors is written%
\begin{eqnarray}
\mathcal{H}_{\rm SC}=\frac{1}{2}\sum_{k}\left(\begin{array}{cc}
                            c_{k}^{\dag}, & c_{-k}
                          \end{array}\right) \left(
\begin{array}{cc}
0 & \Delta _{k} \\
\Delta _{k}^{\dagger} & 0%
\end{array}%
\right) \left(
\begin{array}{c}
c_{k} \\
c_{-k}^{\dagger}%
\end{array}%
\right) .
\end{eqnarray}%
with $\Delta_{\bm{k}}=2\Delta[-\sin k_{y}+i\sin k_{x}]\mathcal{I}$.
After the rotational transformation, it turns to be %
\begin{eqnarray}
\bar{\mathcal{H}}_{\rm SC}=\frac{1}{2}\sum_{k}\left(\begin{array}{cc}
                            \bar{c}_{k}^{\dag}, & \bar{c}_{-k}
                          \end{array}\right) \left(
\begin{array}{cc}
0 & \bar{\Delta}_{k} \\
\bar{\Delta}_{k}^{\dagger} & 0%
\end{array}%
\right) \left(
\begin{array}{c}
\bar{c}_{k} \\
\bar{c}_{-k}^{\dagger}%
\end{array}%
\right)
\end{eqnarray}%
with $\bar{c}_{\bm{k}}^{\dag}=(\alpha_{k}^{\dag}, \beta_{k}^{\dag})$, and
\begin{eqnarray}
\bar{\Delta}_{k}=\left(
\begin{array}{cc}
\Delta _{k}^{\alpha\alpha} & \Delta _{k}^{\alpha\beta} \\
\Delta _{k}^{\beta\alpha} & \Delta _{k}^{\beta\beta} %
\end{array}%
\right).
\end{eqnarray}
More explicitly, $\bar{\mathcal{H}}_{\rm SC}=\sum_{\gamma\eta}\bar{\mathcal{H}}_{\rm SC}^{\gamma\eta}$ ($\gamma$, $\eta=\alpha$, $\beta$) with
\begin{equation}
\bar{\mathcal{H}}_{\rm SC}^{\gamma\eta}=\frac{1}{2}\sum_{k}\left( \Delta _{k}^{\gamma\eta}\gamma_{k}^{\dag }\eta_{-k}^{\dag }+{\rm H.c.}\right).
\end{equation}%
Here, $\Delta _{k}^{\alpha\alpha}= \left( \cos ^{2}\theta
_{k}+e^{-2i\varphi _{k}}\sin ^{2}\theta _{k}\right) \Delta
_{k}^{0}$, $%
\Delta _{k}^{\beta\beta}= \left( \cos ^{2}\theta
_{k}+e^{2i\varphi _{k}}\sin ^{2}\theta _{k}\right) \Delta
_{k}^{0}$, and $%
\Delta _{k}^{\alpha\beta}=\Delta _{k}^{\beta\alpha}=\frac{1}{2}\sin
2\theta _{k}\left( e^{i\varphi _{k}}-e^{-i\varphi _{k}}\right) \Delta
_{k}^{0}$ with $\Delta^{0}_{k}=2\Delta(-\sin k_{y}+i\sin k_{x})$. Therefore, the superconducting terms include both the intra-CI band pairing (%
$\alpha_{k}^{\dagger}\alpha_{-k}^{\dagger}$ or $\beta_{k}^{\dagger}\beta%
_{-k}^{\dagger}$) and inter-CI band pairng ($\alpha_{k}^{\dagger}\beta%
_{-k}^{\dagger}$ or $\beta_{k}^{\dagger}\alpha_{-k}^{\dagger}$) channels.

Now we turn to discuss the particle-hole symmetry in respective pairing channel. The Hamiltonian can be written as
 \begin{eqnarray}
\bar{\mathcal{H}}^{\gamma\eta}=\frac{1}{2}\sum_{k}\left(
\begin{array}{cc}
\gamma_{k}^{\dag } & \eta_{-k}%
\end{array}%
\right) \bar{\mathcal{H}}^{\gamma\eta}(k)\left(
\begin{array}{c}
\gamma _{k} \\
\eta _{-k}^{\dag }%
\end{array}%
\right)
\end{eqnarray} with $\bar{\mathcal{H}}^{\gamma\eta}(k) =\left(
\begin{array}{cc}
\xi _{k}^{\gamma} & \Delta _{k}^{\gamma \eta } \\
\left( \Delta _{k}^{\gamma \eta }\right) ^{\ast } & -\xi _{-k}^{\eta }%
\end{array}%
\right)$. The intra-CI band channel ($\gamma=\eta$), $\mathcal{P}\bar{\mathcal{H}}^{\gamma\gamma}(k)\mathcal{P}^{-1}=-\bar{\mathcal{H}}^{\gamma\gamma}(-k)$, is protected by the particle-hole symmetry. In contrast, the inter-CI band channel ($\gamma\ne\eta$), $\mathcal{P}\bar{\mathcal{H}}^{\gamma\eta}(k)\mathcal{P}^{-1}=-\bar{\mathcal{H}}^{\eta\gamma}(-k)$, is not protected by the particle-hole symmetry. Here, $\mathcal{P}=\sigma_{x}K$ is the operator for the particle-hole transformation with $\sigma$ the $x$-component of Pauli matrix and $K$ the complex conjugation. As well known, the Majorana fermions are their own anti-particle. They naturally reserve the particle-hole symmetry. Therefore, only the intra-CI band pairing contributes the Majorana zero modes. However, this does not mean that the intrinsic particle-hole symmetry is broken in the BdG Hamiltonian since $\mathcal{P}(\bar{\mathcal{H}}^{\gamma\eta}(k)+\bar{\mathcal{H}}^{\beta\gamma}(k))\mathcal{P}^{-1}=-(\bar{\mathcal{H}}^{\gamma\eta}(-k)+\bar{\mathcal{H}}^{\eta\gamma}(-k)$
even when $\gamma\ne\eta$. This can be more evident when we consider the particle-hole symmetry for the whole BdG Hamiltonian. The basis after rotational transformation is arranged as $\psi_{k}=(\alpha_{k},\beta_{k},\alpha_{-k}^{\dag},\beta_{-k}^{\dag})^{T}$. The BdG Hamiltonian is then expressed as
\begin{equation*}
\bar{\mathcal{H}}=\sum_{k}\psi _{k}^{\dag }\bar{\mathcal{H}}\left( k\right) \psi _{k}
\end{equation*}%
with
\begin{equation*}
\bar{\mathcal{H}}\left( k\right) =\left(
\begin{array}{cccc}
\xi _{k}^{\alpha } & 0 & \Delta _{k}^{\alpha \alpha } & \Delta _{k}^{\alpha
\beta } \\
0 & \xi _{k}^{\beta } & \Delta _{k}^{\beta \alpha } & \Delta _{k}^{\beta
\beta } \\
\Delta _{k}^{\alpha \alpha \ast } & \Delta _{k}^{\beta \alpha \ast } & -\xi
_{k}^{\alpha } & 0 \\
\Delta _{k}^{\alpha \beta \ast } & \Delta _{k}^{\beta \beta \ast } & 0 &
-\xi _{k}^{\beta }%
\end{array}%
\right) .
\end{equation*}
The particle-hole symmetry is then protected by $\mathcal{P}\bar{\mathcal{H}}(k)\mathcal{P}^{-1}=\bar{\mathcal{H}}(-k)$, in which the particle-hole transformation operator $\mathcal{P}=\sigma_{x}\otimes\mathcal{I} K$ with $\mathcal{I}$ the $2\times 2$ identity matrix.

\section*{Appendix II: Hamiltonian in cylindrical geometry}
\label{Adix2}
We consider the cylindrical geometry with open boundary condition along $x$-direction and periodic boundary condition along $y$-direction, which is used to calculate the edge states. The Hamiltonian is expressed as
\begin{equation}
\mathcal{H}=\sum_{k_{y}}\left( \psi _{k_{y}}^{\dag },\psi _{-k_{y}}\right) \left(
\begin{array}{cc}
\mathcal{H}_{0}\left( k_{y}\right)  & \Delta \left( k_{y}\right)  \\
\Delta ^{\dag }\left( k_{y}\right)  & -\mathcal{H}_{0}^{T}\left( -k_{y}\right)
\end{array}%
\right) \left(
\begin{array}{c}
\psi _{k_{y}} \\
\psi _{-k_{y}}^{\dag }%
\end{array}%
\right),\label{HKY}
\end{equation}%
where the basis $\psi _{k_{y}}=\left( a_{1,k_{y}},\cdots
,a_{M,k_{y}},b_{1,k_{y}},\cdots ,b_{M,k_{y}}\right) ^{T}$ with $M$ the number of sublattice along the finite direction.
\begin{eqnarray*}
\mathcal{H}_{0}\left( k_{y}\right)&=&(2t^{\prime}\cos k_{y}-\mu)\sum_{m=1}^{M}\left(
a_{m,k_{y}}^{\dag }a_{m,k_{y}}-b_{m,k_{y}}^{\dag }b_{m,k_{y}}\right)  \\
&-&2t\cos\left(\phi+k_{y}/2\right)\sum_{m=1}^{M}\left( a_{m,k_{y}}^{\dag }b_{m,k_{y}}+H.c.\right)\\
&-&2t\cos\left(\phi-k_{y}/2\right)\sum_{m=2}^{M}\left(a_{m,k_{y}}^{\dag }b_{m-1,k_{y}}+H.c.\right)\\
&-&t^{\prime }\sum_{m=1}^{M-1}\left( a_{m,k_{y}}^{\dag }a_{m+1,k_{y}}-b_{m,k_{y}}^{\dag }b_{m+1,k_{y}}+H.c.\right)  \\,
\end{eqnarray*}%
\begin{eqnarray*}
\Delta \left( k_{y}\right)&=&\Delta \sum_{m=1}^{M-1}\left( a_{m,k_{y}}^{\dag
}a_{m+1,-k_{y}}^{\dag }+b_{m,k_{y}}^{\dag }b_{m+1,-k_{y}}^{\dag
}+H.c.\right)  \\
&-&\Delta\sum_{m=2}^{M}\left(a_{m,k_{y}}^{\dag }a_{m-1,-k_{y}}^{\dag }+b_{m,k_{y}}^{\dag }b_{m-1,-k_{y}}^{\dag }+H.c.\right)\\
&-&2\Delta\sin k_{y} \sum_{m=1}^{M}\left(a_{m,k_{y}}^{\dag }a_{m,-k_{y}}^{\dag }+b_{m,k_{y}}^{\dag
}b_{m,-k_{y}}^{\dag }+H.c.\right) .
\end{eqnarray*}
The particle-hole symmetry in Eq.~(\ref{HKY}) is guaranteed by $\mathcal{P}\mathcal{H}(k_{y})\mathcal{P}^{-1}=-\mathcal{H}(-k_{y})$, in which the operator of particle-hole symmetry $\mathcal{P}=\sigma_{x}\otimes \mathcal{I} K$ with $\mathcal{I}$ the $2M\times 2M$ identity matrix and $K$ the complex conjugation. The particle-hole symmetry in cylindrical geometry is evident in upper panels of Fig.~\ref{f.3} in main text, where the band structure seems to be antisymmetric about $k_{y}$. To ensure the zero energy modes, the optional choice is $k_{y}=0$ or $k_{y}=\pi$.

\section*{References}
\bibliography{TSC}

\end{document}